\documentclass[conference]{IEEEtran}
\usepackage{graphicx}
\usepackage[section]{placeins}
\usepackage{xcolor}
\usepackage{colortbl}
\usepackage{lipsum}
\usepackage{listings}
\usepackage{amsmath}
\usepackage{color} 
\usepackage{url}
\definecolor{mygreen}{RGB}{28,172,0} 
\definecolor{mylilas}{RGB}{170,55,241}

\begin{document}

\title{Nonlinear Control of Aerial Manipulator}

\author{\IEEEauthorblockN{Sandesh Thapa}
\IEEEauthorblockA{School of Mechanical and Aerospace Engineering\\
Control, Robotics \&  Automation Lab (CoRAL), Spring 2017 \\
Oklahoma State University\\
Stillwater, Oklahoma\\
Email: sandesh.thapa@okstate.edu}}

\maketitle

\begin{abstract}
This paper presents two controllers for a quadrotor equipped with a two-DOF robot manipulator that physically interacts with the environment. Cartesian impedance and adaptive sliding mode controllers are implemented and compared for a simple hovering task in the presence of some bounded disturbance. A full dynamic model is developed, and the two above controllers are implemented with the same initial conditions and desired position.
\end{abstract}

\IEEEpeerreviewmaketitle

\section{Summary}


\subsection{Cartesian Impedance Control}
As part of the ARCAS [3] project, Lippiello and Ruggiero [1] proposed a nonlinear model-based Cartesian impedance controller in which the nonlinear dynamics of the system are taken into account. Their Cartesian impedance model is a dynamic relationship between the end-effector motion in Cartesian coordinates and the generalized external forces and moments acting on the system. Impedance control is often used in robot manipulation tasks. Classical impedance control requires the measurement of the external force, which typically acts on the manipulator end-effector. So, Lippiello and Ruggiero [1] modified the classical impedance control as suggested by Ott [4], since it is infeasible to measure such an external force in aerial robotics.

\subsubsection{Control Problem}
A UAV equipped with a robotic arm can be an efficient solution for dexterous manipulation. It can also be used for rescue operations, construction in inaccessible areas, and inspection of power lines. However, the presence of a robot arm changes the dynamic behavior of the system and creates coupling effects. The center of mass is no longer constant, and the moment of inertia changes significantly. So, in this paper [1], the focus is on hovering control in the presence of some bounded disturbances.

\subsubsection{Control Law}

\begin{equation}
	\dot{x} = J_a\dot{q}
	\end{equation}
	\begin{equation}
	\ddot{x}= J_a\ddot{q}+ \dot{J_a}\dot{q}
	\end{equation}
	\begin{equation}
	\tilde{x} = x_d - x
	\end{equation}
	\begin{equation}
	\tau = J_a^{T}f
	\end{equation}
	From (1),(2),(3),(3),(4) the joint space dynamics can be transformed into 		    manipulator end-effector Cartesian coordinates (x). Here, $J_a$ is the Jacobian of the system. $		\tau$ is the generalized torque related to the four motor thrusts and joint actuators.\\
	\begin{equation}
	M_x\ddot{x}+ C_x(x,\dot{x})\dot{x} + G_x(x) = f + f_{ext}
	\end{equation}
	where, $M_x = J_a^{-T}M_q J_a^{-1}$, $C_x =  J_a^{-T}(C_q- M_qJ_a^{-1}\dot{J_a})J_a^{-1}$, \\ 		$G_x = J_a^{-T}G_q$,$f= J_a^{-T}\tau$, $f_ext =J_a^{-T}\tau_{ext} $\\
%

The matrices in (5) have the following properties:
\begin{itemize}
\item \textbf{Property 1.} The inertia matrices $M$ and $M_x$ are symmetric and positive definite.\\
\item \textbf{Property 2.} $(M-2C)$ and $(M_x-2C_x)$ are skew-symmetric.
\end{itemize}

The following control law is defined, where $K_P$ and $K_D$ are $n_q\times n_q$
symmetric and positive definite matrices.

\begin{equation}
	\tau = G+ J_a^{T}(M_x\ddot{x_d}+ C_x\dot{x_d} + K_D\dot{\tilde{x}}+K_P\dot{x})
	\end{equation}
	The closed-loop dynamics of the system are:
	\begin{equation}
	M_x\ddot{\tilde{x}}+ (C_x+K_D)\dot{\tilde{x}}+ K_P\tilde{x} = f_{ext}
	\end{equation}

\subsubsection{Stability Analysis}

A positive definite candidate Lyapunov function is defined as:
	\begin{equation}
	V(x,t) = \frac{1}{2}\dot{\tilde{x}}^{T}M_x{\dot{\tilde{x}}} + \frac{1}{2}{\tilde{x}}^{T}	    	K_P{\tilde{x}}
	\end{equation}
	Taking the time derivative of $V(x,t)$:
\begin{equation}\label{eq:10}
\dot{V} = \frac{1}{2}\dot{\tilde{x}}^{T}\dot{M_x}{\tilde{x}} + \dot{\tilde{x}}^{T}M_x\ddot{{\tilde{x}}} + {\tilde{x}}^{T}K_P\dot{{\tilde{x}}}
\end{equation}
substituting equation 7 into 9, we get:

\begin{equation}\label{eq:11}
\dot{V} = - \dot{{\tilde{x}}}^{T}K_D\dot{{\tilde{x}}} + {\dot{\tilde{x}}}^{T}f_{ext}
\end{equation}

	If $f_{ext}= 0$ for the free motion case, the stability analysis of the system is not obvious since the matrix $M_x$ is time-varying, so we cannot use the invariance principle. Hence, only stability can be ensured. Also, it is mentioned in the paper that by considering the Cartesian configuration space in which the Jacobian is invertible, it is possible to show the asymptotic stability of the system. It also states that in the case of constrained motion, $f_{ext}\neq 0$, stability can be ensured only in the unconstrained Cartesian directions.

\section{Critique}
\subsection{Methodology }
The Cartesian impedance model is a model-based controller that needs the exact values of all parameters in the dynamics of the UAV and manipulator. Moreover, in this paper [1], some of the impedance control parameters cannot be specified arbitrarily, and they depend on the parameters of the system dynamics. Also, their controller is focused on the hovering task. For the tracking problem, the authors have not given any results. Very few reports exist on handling the aerial vehicle and the robotic arm as an integrated system. This report [1] was one of the early papers to develop a full dynamic model of a multi-DOF robot arm using Euler-Lagrangian formalism.

\subsection{Literature Review}

The motivation behind introducing this paper was to develop a full dynamic model of the UAV plus robotic arm as a single rigid body. Since the coupling effects and changes in the center of mass and inertia are significant, the dynamic model given seems to be the best way to go forward for this kind of system. Previous reports have controlled the UAV and arm separately, using different controllers for the UAV and the robot arm.

\subsection{Dynamic Model/Mathematics}

This paper gave a complete formulation for the complete dynamics of the UAV plus a robot as a single system. In doing so, they have made many assumptions which might not always hold true. The Jacobian $J_a$ might not always be invertible. Also, the computation of the C matrix becomes very complicated. Their stability analysis is not detailed. In order to prove global asymptotic stability, we need to find a different Lyapunov function for the specific controller mentioned in this paper. Furthermore, they went on to assume that just the controller used for a ground manipulator can be used to prove global asymptotic stability in the UAV plus manipulator situation, which might not always be the case.

Details in the derivation are lacking. While deriving the Jacobian, the authors did not mention any formula for the Jacobian. They also did not give any clear distinction between the manipulator velocity and angular Jacobian. I had to find that using a different paper [2].

\subsection{Simulation Results}

Based on the results provided in the paper, the simulation results show agreement with the theoretical analysis. But they are lacking details on that too. Complete results of torque and generalized force should have been provided in the paper to check the range of torque involved. Any experimental data is lacking, so it is hard to analyze the practical aspects of the controller given.
Their controller is used just for hovering and perturbations. So, the results are not comprehensive. It would have been a better idea if they had provided a plot of the tracking error along with hovering.
\subsection{Suggestions}
I would like to give the following suggestions regarding the paper I used for the simulation:
\begin{itemize}
	\item Use of better notation while deriving the complete dynamics.
	\item Providing better equations for the system Jacobian and manipulator Jacobian.
	\item Use of better labels in the figure while marking the inertial and body frames of the UAV.
	\item Improving the stability analysis.
	\item Include the experimental data.
\end{itemize}

\subsection{Sliding Mode Control}
I used another controller from [2] to compare with the controller mentioned above. \\
\textbf{Control Law:}
\begin{equation}
e = q-q_d
\end{equation}
\begin{equation}
s = \dot{e}-\Lambda e
\end{equation}
\begin{equation}
q_r = \dot{q}-s = \dot{q_d} - \Lambda e
\end{equation}

To design the controller, we define the error $e$ and the sliding surface $s$. Now the control law becomes:

\begin{equation}
\tau = \hat{M}\ddot{q_r}+ \hat{C}\dot{q_r} + \hat{\Delta} -  As - Ksgn(s)
\end{equation}

Here, $\hat{M}$, $ \hat{C}$, $\hat{G}$ represent estimates of each matrix. $A$ and $K$ are positive gain matrices.

\begin{equation}
M\dot{s} + Cs + \Lambda s = -\Delta + \hat\Delta - Ksgn(s)
\end{equation}
where,
\begin{equation}
\Delta = \tilde{M}\ddot{q_r} -\tilde{C}\dot{q_r} - \tilde{G}
\end{equation}

\textbf{Stability Analysis:}
Let a Lyapunov candidate function be
\begin{equation}\label{eq:20}
V = \frac{1}{2}s^TMs + \frac{1}{2}\tilde{\Delta}^T\tilde{\Delta} > 0
\end{equation}

Its time derivative is:

\begin{equation}\label{eq:21}
\dot{V} = \frac{1}{2}s^T\dot{M}s + s^TM\dot{s} +  \tilde{\Delta}^T\dot{\tilde{\Delta}}
\end{equation}
\begin{equation}\label{eq:22}
\dot{V} = \frac{1}{2}s^T(-\Lambda s + \tilde{\Delta} - Ksgn(s) + \tau_{ext}) +  \tilde{\Delta}^T\dot{\tilde{\Delta}} \\
\end{equation}

$\dot{V} = -s^T\Lambda s - S^T(Ksgn(s) - \tau_{ext}) \leq 0$\\
\textbf{Adaptation law: }
\begin{equation}\label{eq:23}
\dot{\hat{\Delta}} - \dot{\Delta} = -s
\end{equation}
with the assumption that $\Delta$ changes very slowly compared with the adaptation rate, we get
\begin{equation}
\dot{\hat{\Delta}} = -s
\end{equation}

Free motion: If we assume that  $\tau_{ext} = 0$, from (19) and (20), we can conclude that $s \to 0$ as t $\to 0$, so $e \to 0$ exponentially. If $\tau_{ext}\neq 0$, we can conclude that s is bounded, so $\implies e $ is bounded.

\section{Simulation Experiment}

		\begin{itemize}
		\item AscTec Pelican, with a mass $m_b$ = 2 kg, inertia matrix $I_b$ = $diag(1.24,1.24,2.48)$
	\item 2-DOF robotic arm, $l_1$ =15 cm $l_2$5 cm with 3 revolute joints,
	mass of link $m_1$ = 0.05 kg, $m_2$ = 0.05 kg, inertia of link $I_1$ = 0.0019 m$^2$kg $I_1$ = 0.0011 m$^2$kg
		\item Initial conditions, $q=\left[0, 0,0,0, 0, 0, 0,0,\frac{\pi}{2}\right]^{T}$
		\item Desired position, $q_d=\left[0, 0,6,0, 0, 0, 0,0,0.5, 0.5\right]$
		\end{itemize}
A force of 1 N is modeled through a sine wave of $\pi/4$ rad/s, which acts along the $x_b$ axis, and another force of 0.5 N magnitude acts along the x-direction of the end-effector of 20 s to model the windy situation. These cases are simulated as done in the paper and compared with another adaptive sliding mode controller.
\subsection{\textbf{Cartesian Impedance Controller}}
\textbf{Case I (Rigid Behavior):} $K_P = 80I_8$, $K_D = 16I_8$
\
	\begin{figure}[!h]
	\centering
		\includegraphics[scale=0.4]{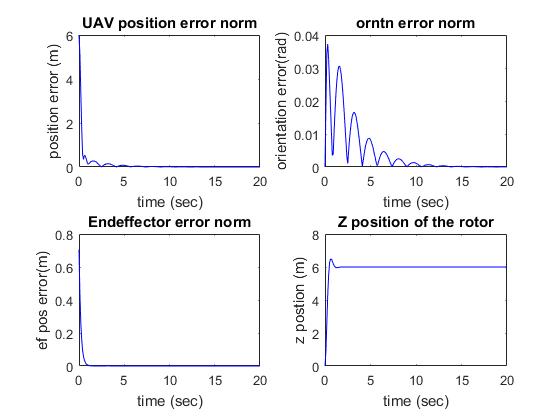}
	\caption{Time histories simulation for Rigid Case.}
	\label{fig}
	\end{figure}

\textbf{Case II (Compliant):} $K_P = 5I_8,$, $K_D = 2I_8$
	\begin{figure}[h!]
	\centering
		\includegraphics[scale=0.35]{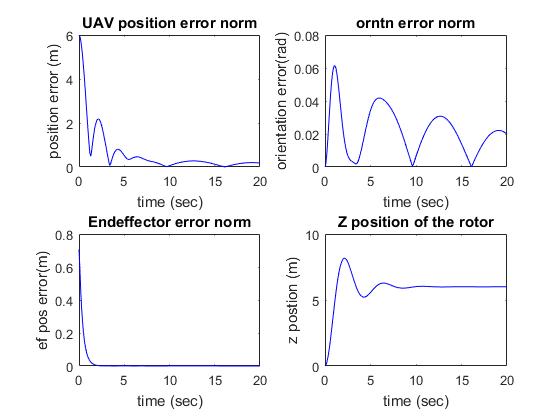}
	\caption{Time histories simulation for Compliant Case.}
	\label{fig}
	\end{figure}

\textbf{Rigid UAV and Compliant manipulator:} $K_P = diag(10I_6,100I_2$), $K_D = diag(3I_6,50I_2)$
	\begin{figure}[h!]
	\centering
		\includegraphics[scale=0.4]{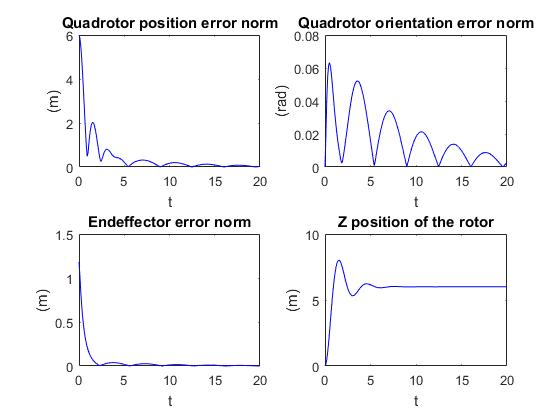}
	\caption{Time histories simulation for Compliant Rigid Case.}
	\label{fig}
	\end{figure}
	\FloatBarrier
\subsection{\textbf{Adaptive Sliding Mode Controller:}}

	Design Parameters: $ K = 8I_{8\times8}$, $ \Lambda = 2I_{8\times8}$ , $ A = 10I_{8\times8}$
	\begin{figure}[!h]
	\centering
		\includegraphics[scale=0.4]{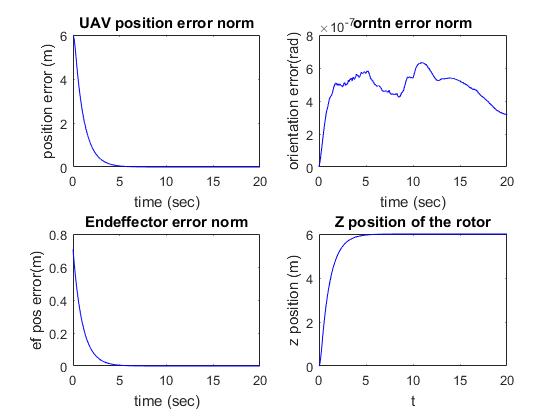}
	\caption{Time histories simulation for adaptive sliding mode controller.}
	\label{fig}
	\end{figure}
	\FloatBarrier
\textbf{Position History}
	\begin{figure}[!h]
	\centering
		\includegraphics[scale=0.4]{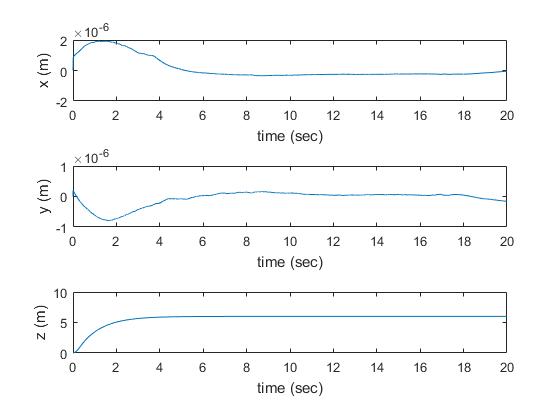}
	\caption{Position history of the quadrotor.}
	\label{fig}
	\end{figure}
	\FloatBarrier

\textbf{Attitude History}
	\begin{figure}[!h]
	\centering
		\includegraphics[scale=0.4]{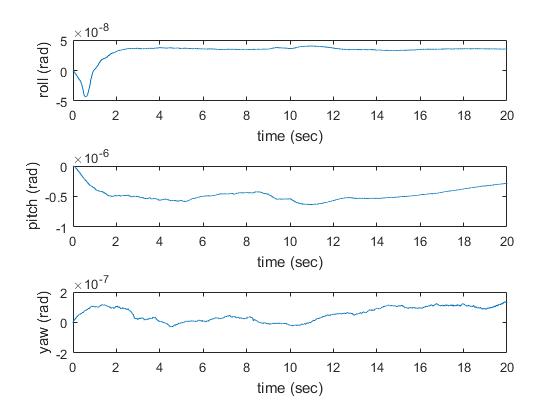}
	\caption{Attitude history of the quadrotor.}
	\label{fig}
	\end{figure}
	\FloatBarrier

\textbf{Torque Plots}

\begin{figure}[!h]
	\centering
		\includegraphics[scale=0.25]{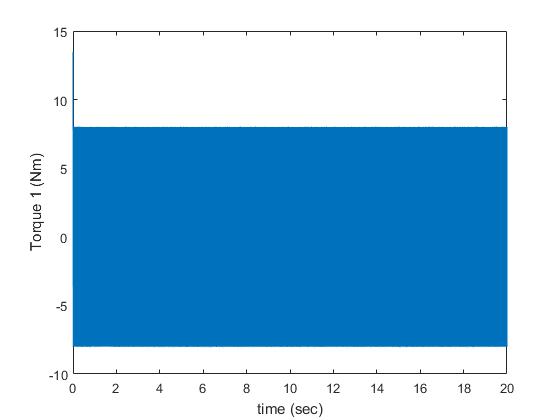}
	\caption{Torque History for link 1}
	\label{fig}
	\end{figure}

\begin{figure}[!h]
	\centering
		\includegraphics[scale=0.25]{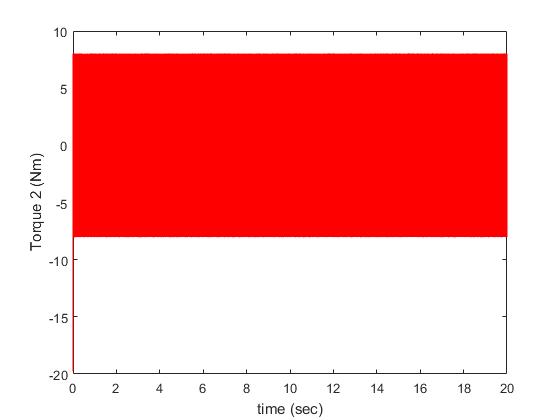}
	\caption{Torque History for link 2}
	\label{fig}
	\end{figure}
	\FloatBarrier

\begin{table}[!h]
  \centering
  \caption{Comparison of response for z-position}
    \begin{tabular}{|c|c|c|c|c|}
    \textbf{Z position}  & \textbf{Rigid}  & \textbf{Compliant}  &\textbf{ CompRigid} & \textbf{AdpSMC}  \\
    overshoot  & 8\%   & 36.05\% & 33.66 & 0 \\
    rise time (sec) & 0.8   & 1.8   & 1.5   & 4.4 \\
    settling time (sec) & 1.8   & 7     & 5     & 5 \\

    \end{tabular}%
  \end{table}

  \FloatBarrier

 \begin{table}[!t]
  \centering
  \caption{Comparison of RMS error for two controllers}
    \begin{tabular}{|c|c|c|c|c|}
    error &position (m) & orientation (rad) & end eff pos (m) \\
    \textbf{Rigid}  & 0.6265 & 0.0088 & 0.0628 \\
    \textbf{Compliant} & 1.2006 & 0.0258 & 0.0811 \\
    \textbf{CompRigid} & 1.0148 & 0.023 & 0.0879 \\
    \textbf{AdpSMC} & 1.0769 & 4.90E-07 & 0.1128 \\
    \end{tabular}%
\end{table}%
\FloatBarrier

\begin{figure}[!h]
	\centering
		\includegraphics[scale=0.35]{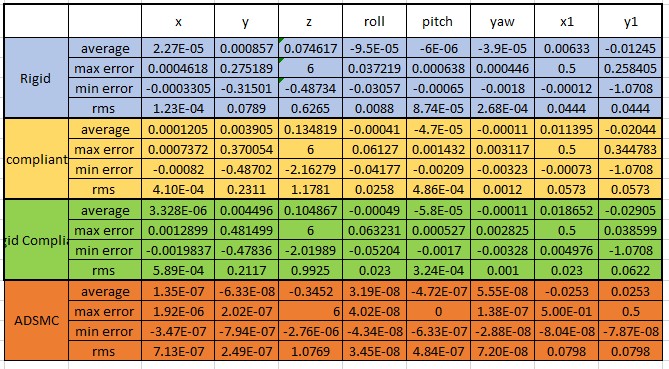}
	\caption{Error comparison for both of the controllers}
	\label{fig}
	\end{figure}
	\FloatBarrier

The simulation results of the Cartesian impedance controller match those in the paper. However, their desired trajectory is not given, so I used the above-stated desired conditions for my simulation purposes. Comparing my plot with the one from the paper, it shows good agreement.

We can observe from figures 1--3 that the error is bounded. The orientation error still will not go to zero because of the external disturbance in the system. The z-position converges from 0 to 6 in less than 5 seconds in all three cases. Also, in the adaptive controller, the error remains bounded, as we can observe from figures 4, 5 \& 6.

While simulating the Cartesian impedance controller, three cases were done in order to find the rigid behavior of the manipulator and the compliant behavior of the UAV. One of the major challenges while implementing the controller was tuning the gain matrices. Comparing the Cartesian controller with the adaptive sliding mode controller, it can be clearly seen from just the z-position history that the adaptive controller is better in terms of performance. There is no overshoot with the adaptive controller. However, in all three cases for the impedance controller, there was significant overshoot. The adaptive controller has a better response but a higher error in the end-effector position. From figure 9, we can observe that the error is significantly less in the adaptive controller except for the end-effector position.

\section{Conclusion}
The coupled dynamics of an aerial manipulator make the control problem challenging. This paper contributed towards the dynamic modeling of an aerial manipulator. Cartesian impedance control is used widely in robotics. However, its relevance in aerial robotics has still not been properly analyzed. Comparing all the results, we can see that the Cartesian impedance controller is better in task space, but in other cases like the position and orientation of the UAV, the adaptive controller is better. So, the solution might be a combination of the two, using an adaptive Cartesian impedance controller.

The MATLAB/Simulink code for the simulations is available at \url{https://github.com/sandeshthapa/Cartesian_Impedance_Control_Aerial_Manipulator} (Cartesian impedance controller) and \url{https://github.com/sandeshthapa/Adaptive_Sliding_Mode_Control_of_Aerial_Manipulator} (adaptive sliding mode controller).

\section{References}
[1] V. Lippiello and F. Ruggiero, ”Cartesian impedance control of a UAV
with a robotic arm,” in 10th International IFAC Symposium on Robot
Control, Dubrovnik, Croatia, Sep 2012.

[2] S. Kim, S. Choi, and H. J. Kim, “Aerial manipulation using a quadrotor
with a two DOF robotic arm,” in Proc. IEEE/RSJ Int. Conf. Intell. Robots
Syst., Nov. 2013, pp. 4990–4995.

[3] The ARCAS (Aerial Robotics Cooperative Assembly system)
project. http://www.arcas-project.eu/.

[4] Ott, C, Cartesian Impedance Control of Redundant and Flexible-Joint Robotics, volume 49 Springer Tracts in Advanced Robotics, 2008, London, UK.

[4]R. M. Murray, Z.Li, and S. S. Sastry, A Mathematical Introduction to
Robotic Manipulation. Boca Raton, FL: CRC, 1994.

[5] Mojtaba Sharifi \& Hassan Sayyaadi (2015) Nonlinear robust adaptive Cartesian
impedance control of UAVs equipped with a robot manipulator, Advanced Robotics, 29:3, 171-186,

[6] Lippiello V, Ruggiero F. Exploiting redundancy in Cartesian
impedance control of UAVs equipped with a robotic
arm. In: 2012 IEEE/RSJ International Conference on
Intelligent Robots and Systems (IROS); 2012 Oct 7–12;
Vilamoura, Portugal. p. 3768–3773

%
%
%

\end{document}